\documentclass[epj]{svjour}
\usepackage{graphicx}
\begin{document}
\title{Parity Violation in Astrophysics}
\author{C. J. Horowitz} 
\institute{Nuclear Theory Center and Department of Physics, Indiana University, Bloomington, IN 47405, USA, e-mail: horowit@indiana.edu}

\date{Received:}

\abstract{
Core collapse supernovae are gigantic explosions of massive stars that radiate 99\% of their energy in neutrinos.  This provides a unique opportunity for large scale parity or charge conjugation violation.  Parity violation in a strong magnetic field could lead to an asymmetry in the neutrino radiation and recoil of the newly formed neutron star.  Charge conjugation violation in the neutrino-nucleon interaction reduces the ratio of neutrons to protons in the neutrino driven wind above the neutron star.  This is a problem for r-process nucleosynthesis in this wind.  On earth, parity violation is an excellent probe of neutrons because the weak charge of a neutron is much larger than that of a proton.  The Parity Radius Experiment (PREX) at Jefferson Laboratory aims to precisely measure the neutron radius of $^{208}$Pb with parity violating elastic electron scattering.  This has many implications for astrophysics, including the structure of neutron stars, and for atomic parity nonconservation experiments.}


\maketitle
\section{Introduction}
\label{intro}
Can there be large scale parity violation in astrophysics?  For example, the 1960s science fiction movie {\it Journey to the Far Side of the Sun} visits a parity double of earth.  In the real world, core collapse supernovae provide a unique opportunity for macroscopic parity violation.  These gigantic explosions of massive stars are dominated by neutrinos that transport 99\% of the energy.  This is because no other known particle can diffuse through the very dense matter of the collapsed star.  In Section \ref{supernovae} we discuss how the weak interactions of these neutrinos may lead to large scale parity or charge conjugation violation.     

On earth, parity violation can be used to probe neutrons because the weak charge of a neutron is much larger than that of a proton.  In Section \ref{PREX} we describe the Parity Radius Experiment (PREX) at Jefferson Laboratory that aims to use parity violating electron scattering to measure the neutron radius in $^{208}$Pb.  This measurement has many implications for astrophysics including properties of neutron stars, nuclear structure, and atomic parity nonconservation.

\section{Parity Violation in Supernovae}
\label{supernovae}
In this section we explore large scale parity or charge conjugation violation in core collapse supernovae.  When the center of a massive star has burned to $^{56}$Fe, thermonuclear reactions cease and the core of the star collapses, in a fraction of a second, all the way to nuclear densities.  This forms a proto-neutron star with half again the mass of the sun and a radius of order 10 kilometers.  The proto-neutron star is initially hot and lepton rich.  Over a period of seconds, its very large gravitational binding energy of over 100 MeV/nucleon is radiated away in neutrinos.  Note that, any electromagnetic or strongly interacting particles diffuse very slowly at these high densities.  This immense neutrino burst involves 10$^{58}$ neutrinos and carries about $3\times 10^{53}$ ergs.  In an historic first, about 20 neutrinos were detected from SN1987A.

\begin{figure}
\resizebox{0.4\textwidth}{!}{%
  \hskip .5 in\includegraphics{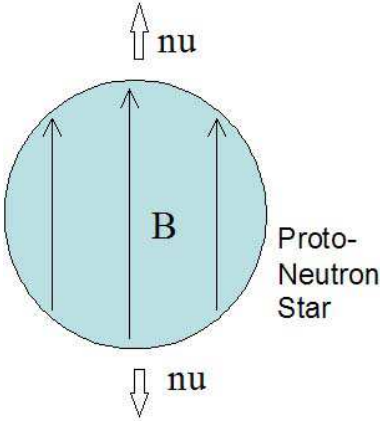}
}
\caption{Parity violation in a strong magnetic field could lead to an up/ down asymmetry in the neutrino flux.}
\label{fig1}      
\end{figure}

The proto-neutron star is so dense that even weakly interacting neutrinos diffuse.  They interact thousands of times before leaving the star.  Could there be a signature of all of these weak interactions?  One possibility is parity violation in a strong magnetic field.  This may lead to an asymmetry in the number of neutrinos emitted along the field compared to those emitted against the field, see Fig. \ref{fig1}.  The neutrinos carry away so much momentum that a few \% asymmetry in the neutrino radiation can lead to a recoil of the neutron star at hundreds of kilometers per second \cite{PVSN}.  Indeed, many neutron stars are observed to have large galactic velocities that presumably came from kicks at birth of 500 km/s or larger.  Unfortunately, explicit calculations of the neutrino asymmetry from parity violation have proved complicated and uncertain.  A very strong magnetic field, of order $10^{15}$ gauss or more, may be required to produce the few percent asymmetry that can explain the observed velocities of neutron stars.

Charge conjugation (C) violation is closely related to parity (P) violation.  In the standard model, the product CP is approximately conserved so the large P violation implies large C violation.  As a result of C violation, neutrino nucleon cross sections are systematically larger than antineutrino-nucleon cross sections.  This changes the composition of the neutrino driven wind during a supernova and may be important for nucleosynthesis.  The intense neutrino flux blows some baryons off of the neutron star and into a high entropy wind.  As this wind cools the nucleons can condense to form heavy nuclei.  This wind is the most promising site for r-process nucelosynthesis.  In the r-process, seed nuclei rapidly capture several neutrons to form about half of the elements heavier than Fe.       

The ratio of neutrons to protons in the wind is set by the rates of neutrino and antineutrino capture.
\begin{equation}
\nu_e + n \rightarrow p + e^-
\label{nucap}
\end{equation}
\begin{equation}
\bar\nu_e + p \rightarrow n + e^+
\label{nubarcap}
\end{equation}
Because of C violation, the cross section for Eq. \ref{nubarcap} is significantly smaller than the cross section for Eq. \ref{nucap}.  This decreases the ratio of neutrons to protons in the wind by 20\%.  As a result, the wind is unlikely to be significantly neutron rich \cite{rprocess} and present simulations do not have enough neutrons to synthesize all of the r-process elements.  Furthermore, it appears very difficult to change conditions to make the wind significantly neutron rich.  This is an important problem for r-process nucleonsynthesis in the neutrino driven wind.  Possible alternative sites, although these also have problems, include neutron star mergers and accretion disks around black holes.  The site of the r-process remains an important open problem in nuclear astrophysics.    

\section{The Parity Radius Experiment (PREX)}
\label{PREX}
Parity violation is a uniquely clean probe of neutron densities.  This is because the weak charge of a neutron is much larger than that of a proton.  Determining neutron densities of heavy nuclei has important implications for the neutron rich matter that is present in many astrophysical objects.  In this section we describe the Parity Radius Experiment (PREX) at Jefferson Laboratory to measure parity violation in elastic electron scattering from $^{208}$Pb, and mention some of its implications.

The parity violating asymmetry $A$ for elastic electron scattering provides a purely electroweak probe of neutron densities.  In Born approximation, $A$ is
\begin{equation}
A={G_F Q^2\over 4\pi 2^{1/2} \alpha} F_W(Q^2)/ F_{ch}(Q^2),
\label{born}
\end{equation} 
and involves the ratio of the weak form factor $F_W(Q^2)$ to the charge form factor $F_{ch}(Q^2)$.  The charge form factor is just the Fourier transform of the charge density at the momentum transfer $Q$ of the experiment and is known from electron scattering.  Likewise the weak form factor is the Fourier transform of the weak charge density.  This directly gives the neutron density since most of the weak charge is carried by neutrons.  At low momentum transfers $A$ is very sensitive to the difference in neutron and proton root mean square radii.

A heavy nucleus such as $^{208}$Pb is expected to have a neutron rich skin because of the neutron excess and because of the Coulomb barrier that prevents protons from being at large distances.  The thickness of this skin has many implications for the properties of neutron rich matter as we discuss below.  However the skin thickness has never been cleanly measured because all hadronic probes of neutron densities have significant uncertainties from the strong interaction.  Parity violation provides a very clean way to measure the neutron skin thickness.    

The Paritiy Radius Experiment (PREX) is proposed at Jefferson Laboratory to measure $A$ for elastic scattering of 850 MeV electrons from $^{208}$Pb at a scattering angle of six degrees \cite{prex}.  The goal is to determine $A$ to $\pm 3\%$.  This allows one to determine the neutron rms radius to 1\% ($\pm 0.05$ fm) \cite{bignradius}.  Because it is a purely electro-weak reaction, the theoretical interpretation of the results is very clean. 

\subsection{Coulomb Distortions and Transverse Analyzing Power}

A heavy nucleus has a large charge $Z$ that will significantly distort the electron waves.  This invalidates the simple Born approximation formula in Eq. \ref{born}.  However, Coulomb distortions can be included exactly by numerically solving the Dirac equation for an electron moving in the Coulomb potential plus the very small (axial-vector) weak potential\cite{distortions}.  This involves summing over a very large number of partial waves, because of the long range of the Coulomb potential.  However, there are standard techniques for improving the convergence of this sum.   We find that Coulomb distortions reduce $A$ by about 30\% \cite{distortions}.  Since the experimental goal is a 3\% measurement it is very important that distortions are accurately included.   

One way to check the distortions is to look at the parity allowed transverse analyzing power $A_y$.  If the initial electron spin is transverse to the scattering plane, the analyzing power $A_y$ gives the asymmetry for scattering to the left compared to scattering to the right.  Time reversal invariance requires $A_y$ to vanish in Born approximation.  However, $A_y$ is nonzero when Coulomb distortions are included.  Therefore, $A_y$ provides a direct test of the same Coulomb distortions.  Furthermore, a nonzero $A_y$ could be an important source of systematic errors.  This is because the beam may have a small residual transverse polarization, and this coupled with $A_y$ could lead to a false asymmetry that changes sign with the beam helicity.  There is considerable interest in $A_y$ for electron-proton scattering \cite{ayp}.  

\begin{figure}
\resizebox{0.4\textwidth}{!}{%
  \hskip .5 in\includegraphics[height=0.99\hsize,angle=-90]{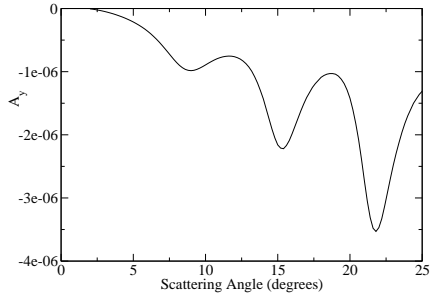}
}
\caption{Analyzing power $A_y$ for elastic scattering of 850 MeV electrons from $^{208}$Pb versus laboratory scattering angle. }
\label{fig2}      
\end{figure}

In Fig. \ref{fig2} we show the transverse asymmetry $A_y$ for the scattering of 850 MeV electrons from $^{208}$Pb.  This is based on an exact numerical solution of the Dirac equation that sums up any number of photon exchanges.  The nucleus is assumed to remain in its ground state at all times.  Intermediate states involving excited states of Pb are not included.  However, the elastic scattering is coherent and leads to a cross section proportional to $Z^2$.  Each intermediate state, involving the excitation of a given proton, is not coherent and contributes a cross section of order unity.  Even when one sums over the $Z$ protons this only gives a total contribution of order $Z$ and this is small compared to the $Z^2$ coherent contribution.  Therefore our elastic approximation should be good for a heavy nucleus like $^{208}$Pb.  We find that $A_y$ is relatively large, comparable in magnitude to $A$ and has significant structure in diffraction minima.  It would be useful to measure $A_y$ using a modest amount of running with transverse polarized beam.  This can be done during PREX.  Note that, $A_y$ grows with $Z$ because it is sensitive to Coulomb distortions.  Therefore $A_y$ is smaller for lighter nuclei.  We will present more $A_y$ results in a future publication \cite{cooper}.

\subsection{Atomic Parity Nonconservation and PREX}
\label{atomic}
Atomic parity nonconservation experiments provide important low energy tests of the standard model.  Parity violation involves the overlap of atomic electrons with the weak charge density of the nucleus and this is primarily carried by the neutrons.  Therefore, precession atomic experiments are sensitive to the neutron radius.  In the future, the most accurate low energy parity test of the standard model may involve the combination of an atomic experiment and PREX to constrain the neutron density\cite{bignradius}. Note that, the Colorado Cs experiment is presently limited by atomic theory uncertainties in the electron density at the nucleus \cite{cs}.  One way to reduce these electron wave function uncertainties is to measure the ratio of parity violation in two different isotopes.  Such a ratio measurement can be very sensitive to the neutron radii of the isotopes.  These radii can be constrained by PREX or future parity violating electron scattering experiments.   

\subsection{The Equation of State of Neutron Rich Matter, PREX and Neutron Stars}
\label{eos}
Measuring the neutron radius in Pb constrains the Equation of State (EOS), pressure versus density, of neutron rich matter.  The pressure forces neutrons, in the skin of a heavy nucleus, out against surface tension.  Therefore, the higher the pressure, the larger will be the neutron radius.  Thus a measurement of the neutron radius in $^{208}$Pb determines the pressure of neutron matter at just below normal nuclear density\cite{brown}.  

A neutron star is a gigantic nucleus, see for example \cite{science} for an introduction.  Its structure depends only on the equations of General Relativity and the EOS of neutron rich matter.  The central density of a neutron star can be a few or more times nuclear density.  Therefore, measuring the radius of a neutron star, which is expected to be of order 10 km, will determine the EOS at high densities \cite{smallNS}.  Astronomers are working hard to measure neutron star radii.  One approach is to determine both the X-ray luminosity and surface temperature.  Thermodynamics allows one to determine the surface area assuming black body radiation plus model corrections for non-black body effects in the atmosphere \cite{NSradii}.    

It is very interesting to compare the low density information on the EOS from the Pb radius with the high density information from a neutron star radius.  There are many possible exotic phases of QCD at high densities.  These include quark matter, strange matter, pion or kaon condensates, and color superconductivity.  An exotic phase could show up as an abrupt softening of the EOS with increasing density.  Note, if an exotic phase has a stiff EOS (high pressure) then it will not be thermodynamicly favored. 

The following scenario would provide an exciting indication of an exotic high density phase.  PREX could measure a large Pb radius.  This shows that the EOS is stiff at low density.   If the radius of a neutron star then turns out to be small, this shows that the high density EOS is relatively soft.  Although these measurements would not, by themselves, determine the nature of the high density phase, they would strongly suggest interesting behavior.  It has proved very difficult to find other signatures of an exotic phase in the center of neutron stars.  Finally, we note that PREX has many other implications for the solid crust of neutron stars \cite{crust} and for how neutron stars cool \cite{urca}.    

\subsection{Neutron Rich Nuclei and PREX}
\label{neutron_rich}
PREX also has important implications for the structure of neutron rich nuclei that can be studied in radioactive beams.  The Rare Isotope Accelerator (RIA) would produce such nuclei and their properties are interesting for several reasons.  For example as discussed in section \ref{supernovae}, r-process nucleosynthesis involves the capture of many neutrons to produce very neutron rich nuclei.

The radius of Pb determines the density dependence of the symmetry energy $S(\rho)$.  The symmetry energy, describes how the energy of nuclear matter rises as one moves away from equal numbers of neutrons and protons. It is very important for the structure of neutron rich nuclei.  The energy of pure neutron matter at density $\rho$, $E_{neutron}(\rho)$ is approximately,
\begin{equation}
E_{neutron}(\rho)\approx E_{nuclear}(\rho) + S(\rho),
\end{equation} 
here $E_{nuclear}$ is the energy of symmetric nuclear matter $(N=Z)$ and is largely known.  Determining the pressure of neutron matter, by measuring the neutron radius, gives $dE_{neutron}/d\rho$ and this gives $dS/d\rho$.  Thus PREX will determine the density dependence of the symmetry energy and this is poorly constrained at present.

\section{Conclusions: Parity Violation and Astrophysics}
\label{conclusions}

Core collapse supernovae are dominated by neutrinos.  This provides a unique opportunity for large scale parity or charge conjugation violation.  Parity violation in a strong magnetic field could lead to an asymmetry in the neutrino radiation and recoil of the newly formed neutron star.  Charge conjugation violation in the neutrino driven wind above a neutron star reduces the ratio of neutrons to protons.  This is a large problem for r-process nucleosynthesis in the wind.

On earth, parity violation probes neutrons because the weak charge of a neutron is much larger than that of a proton.  The Parity Radius Experiment (PREX) at Jefferson Laboratory aims to measure the neutron radius of $^{208}$Pb to 1\% by using parity violating elastic electron scattering.  Because it is an electroweak reaction, the theoretical interpretation of this measurement is very clean.  The neutron radius of Pb has many implications for atomic parity nonconservation experiments, neutron stars, and the structure of neutron rich nuclei.   

\section{Acknowledgments}
Some of this work was done in collaboration with Jorge Piekarewicz.  We thank E. D. Cooper for help on the $A_y$ calculations.  This work was supported in part by DOE grant DE-FG02-87ER40365.


\begin{thebibliography}{}
\bibitem{PVSN}C. J. Horowitz and J. Piekarewicz, Nucl.Phys. {\bf A640} (1998) 281. 
\bibitem{rprocess}C. J. Horowitz, Phys.Rev. {\bf D65} (2002) 083005. 
\bibitem{prex}Jefferson Laboratory Experiment E-03-011, P. A. Souder, R. Michaels and G. Urciuoli spokespersons.  See also http://hallaweb.jlab.org/parity/prex 
\bibitem{bignradius}C. J. Horowitz, S. J. Pollock, P. A. Souder and R. Michaels, Phys.Rev. {\bf C63} (2001) 025501.  See also http://cecelia.physics.indiana.edu/prex  
\bibitem{distortions} C. J. Horowitz, Phys.Rev. {\bf C57} (1998) 3430.
\bibitem{ayp} F. E. Maas et al., nucl-ex/0410013. 
\bibitem{cooper}E. D. Cooper and C. J. Horowitz, to be published.
\bibitem{cs} Wood CS, Bennett SC, Cho D, Masterson BP, Roberts JL, Tanner CE, Wieman CE, Science {\bf 5307} (1997) 1759.
\bibitem{brown} B. A. Brown, Phys. Rev. Lett., {\bf 85} (2000) 5296.
\bibitem{science}J. M. Lattimer and M. Prakash, Science {\bf 304} (2004)536.
\bibitem{smallNS} J. Carriere, C. J. Horowitz, and J. Piekarewicz, Astrophys.J. {\bf 593} (2003) 463.
\bibitem{NSradii}Jose Pons et al., Astrophys. J. {\bf 564} (2002) 981.
\bibitem{crust}C. J. Horowitz and J. Piekarewicz, Phys.Rev.Lett. {\bf 86} (2001) 5647.
\bibitem{urca} C. J. Horowitz and J. Piekarewicz, Phys.Rev. {\bf C66} (2002) 055803.
\end{thebibliography}
\end{document}